\documentclass[%
a4paper,
noarxiv,
onecolumn,
]{quantumarticle}

\makeatletter
\disable@package@load{lmodern}{}
\makeatother

\usepackage{amssymb}

\usepackage{xltxtra}

\usepackage{polyglossia}

\usepackage{xecyr}

\setdefaultlanguage{english}
\setotherlanguages{russian}

\usepackage{fontspec}
\defaultfontfeatures{Scale=MatchLowercase} 

\usepackage[%
  DefaultFeatures={Scale=0.94},
  TT={Scale=MatchLowercase,FakeStretch=0.9},
  DefaultFeatures={Ligatures=Common},
]{plex-otf}

\addfontfeatures{Fractions=On}

\usepackage{unicode-math}

\unimathsetup{math-style=ISO}

\usepackage[%
  parentracker=true,%
  backend=biber,%
  hyperref=auto,%
  language=auto,%
  autolang=other*,%
  defernumbers=true,%
  style=nature,%
  url=false,
  isbn=false,
  doi=true,
  eprint=true,
  maxcitenames=3,
  maxbibnames=100,
  maxnames=10,%
  minnames=1,%
  sorting=none,
  articletitle=true,%
  intitle=true,%
]{biblatex}

\usepackage{amsmath}
\usepackage{amssymb}

\usepackage{mathtools}
\mathtoolsset{
showonlyrefs,
mathic = true
}

\allowdisplaybreaks

\usepackage{hyperref}
\hypersetup{backref,
 colorlinks=false}
\hypersetup{pdfborder=0 0 0}

\usepackage{microtype}
\UseMicrotypeSet[protrusion]{alltext}

\usepackage{graphicx}

\usepackage[scanall]{psfrag}

\usepackage{listings}
\usepackage{listingsutf8}
\usepackage[frozencache]{minted}

\setminted{breaklines=true}

\usepackage{nicefrac}

\makeatletter
\def\ps@pprintTitle{%
     \let\@oddhead\@empty
     \let\@evenhead\@empty
     \let\@oddfoot\@empty
     \let\@evenfoot\@oddfoot}
\makeatother

\begin{document}

\graphicspath{{image/reproducible-network-research/en/}{image/reproducible-network-research/}{image/}}

\title{Reproducible Research in Network Modeling}

\author{Anna M. Ermolayeva}
\email{ermolaevaanna@bk.ru}
\affiliation{RUDN University, 6 Miklukho-Maklaya St, Moscow, 117198, Russian Federation}

\author{Tatyana R. Velieva}
\email{velieva-tr@rudn.ru}
\affiliation{RUDN University, 6 Miklukho-Maklaya St, Moscow, 117198, Russian Federation}

\author{Anna A. Zhivtsova}
\email{zhivtsova-aa@rudn.ru}
\affiliation{RUDN University, 6 Miklukho-Maklaya St, Moscow, 117198, Russian Federation}

\author{Anna V. Korolkova}
\email{korolkova-av@rudn.ru}
\affiliation{RUDN University, 6 Miklukho-Maklaya St, Moscow, 117198, Russian Federation}

\author{Dmitry S. Kulyabov}
\email{kulyabov-ds@rudn.ru}
\affiliation{RUDN University, 6 Miklukho-Maklaya St, Moscow, 117198, Russian Federation}
\affiliation{Joint Institute for Nuclear Research, 6 Joliot-Curie St, Dubna, 141980, Russian Federation}

    \begin{abstract}
        \emph{Background}
        When we model networks, there is a problem of obtaining experimental data to verify other model approaches.
        And even if there are some experimental data, it is necessary to be sure of their reliability.
        \emph{Purpose}
        It is necessary to propose methods for obtaining reliable experimental data.
        \emph{Method}
        By its nature, network equipment is a software and hardware complex.
        Therefore, a full-scale software model can be considered completely equivalent to real equipment.
        And a real experiment can be replaced by a nature experiment.
        The reliability of a nature experiment will be based on its reproducibility.
        \emph{Results}
        A comparison of popular nature network modeling packages was carried out.
        These packages were divided by functionality and feasibility of reproducible studies.
        \emph{Conclusions}
        Most software packages meet the reproducibility criteria.
        The choice of a specific solution depends on non-technical factors: popularity and knowledge of the package.
    \end{abstract}

    \keywords{%
        natural modeling,
        reproducible research,
        research as code
    }

        \maketitle

    \section{Introduction}
\label{sec:intro}

    During modeling networks, there is a problem of access to experimental data.
    It is rare that there is an opportunity to conduct an experiment on real equipment.
    Without experimental data, we lose the ability to verify the results of different types of modeling.
    To remove this problem, it is recommended to conduct a natural experiment using a natural modeling tool.
    It should be noted that natural modeling of computer systems is distinguished by the fact that computer systems can be accurately represented using other computer systems.
    However, this raises the problem of reproducibility of the natural experiment.
    The purpose of the article is to select a natulal modeling tool when the focus is on the support of this tool for conducting a reproducible experiment.

    \subsection{Structure of the paper}
\label{sec:structure}

    Section \ref{sec:model} provides a brief overview of modeling methods.
    Section \ref{sec:reproducible} discusses the problems that hinder reproducible research.
    Section \ref{sec:soft} provides an overview of tools for natural network modeling.
    The authors have tried these modeling tools and described their impressions of them.
    Section \ref{sec:results} summarizes the characteristics of the considered packages related to reproducible calculations.
    Section \ref{sec:discussion} discusses which packages are most suitable for reproducible natural modeling.
    Section \ref{sec:conclusion} provides our specific recommendations for using the considered modeling packages.

	\section{Model approaches}
\label{sec:model}

\begin{figure}
  \centering
  \includegraphics[width=0.6\linewidth]{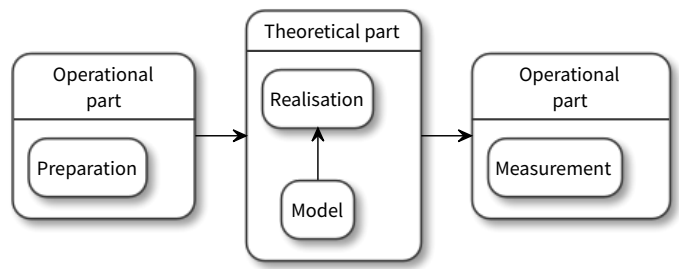}
		\caption{General structure of the model approach}
  \label{fig:model-generic}
\end{figure}

	Modeling as a discipline covers different types of model approaches~\cite{kulyabov_2024_dcm_editorial_journal-rubrics_en}.
	From our point of view, these approaches can be schematically described in a unified way.
	The research structure consists of operational and theoretical parts (Fig.~\ref{fig:model-generic}).
	The operational parts are represented by the procedures for preparing the system and measuring. It is also possible to describe the operational parts as input and output data.
	The theoretical part consists of two layers: the model layer and the implementation layer.

	The model layer is the main one and defines the model being studied.
	The implementation layer describes the specific structure of the system evolution. Depending on the type of implementation, different types of models can be obtained:
	\begin{itemize}
		\item implementation---mathematical expressions: analytical mathematical models;

		\item implementation---analog system: physical model;

		\item implementation---equivalent system: natural model;

		\item implementation---algorithm: simulation models;

		\item implementation---behavior approximation: surrogate model.
	\end{itemize}

	Each type of model has its own area of applicability, its own advantages and disadvantages.
	Using the entire spectrum of models allows for the most in-depth and comprehensive study of the modeled system.

	\subsection{Analytical modeling}

	The most rigorous study is usually based on an analytical mathematical model.
	In this case, the model layer is implemented through mathematical expressions describing the evolution of the system.

	\subsection{Physical modeling}

	The resulting mathematical model must be compared with experimental data and verified.
	To do this, you can create a model similar to a physical or technical system.
	The study is conducted on installations that preserve the nature of the phenomena, but are not necessarily identical to the real object.
	The model can be a reduced or enlarged copy of the prototype, where compliance with the similarity criteria is critical.
	Physical modeling is based on similarity theory and dimensional analysis.

	\subsection{Natural modeling}

	Experiments are conducted on a real object or part of it under conditions as close as possible to operational ones.
	Includes production tests, scientific experiments with intervention in the process and analysis of accumulated experience.
	A natural model can also be virtual.
	For example, you can build a model of a data transmission network using images of operating systems of routers and switches.
	Using a virtual model radically simplifies the modeling process.

	\subsection{Simulation modeling}

	With the development of computer technology, it became possible to specify a model implementation not in the form of a mathematical description, but in the form of some algorithm.
	This type of model is called a simulation model, and the approach itself is called simulation modeling.
	The simulation model plays a dual role.
	A well-established and tested simulation model on experimental data, physical and natural models can in itself serve the purpose of verifying a mathematical model.
	On the other hand, a simulation model allows more effectively than a mathematical model to study the behavior of the modeled system with different variants of input data.

	\subsection{Statistical modeling}

	This is a completely separate direction, which uses correlation rather than causation for automatic model building.
	This type of modeling includes models that are implemented using machine learning methods.
	It can be divided into several approaches.
	For the purposes of modeling itself, surrogate modeling is most interesting.
	In this approach, the model layer is known and even has an implementation (usually in the form of an analytical or simulation model).
	The original modeling can take quite a long time.
	To simplify the study, models are built that imitate the behavior of the original model as closely as possible, while being computationally cheap.
	Surrogate models are built using a data-driven approach.
	The exact internal workings of the modeling code are not assumed to be known (or even understood), only the input-output (preparation-measurement) behavior is important.
	The scientific goal of surrogate modeling is to create a surrogate that is as accurate as possible using as few modeling estimates as possible~\cite{kulyabov_2019_ceur-ws_2507_structural-deep-learning_en}.

	\subsection{Multi-model approach}

	When modeling, it is natural to use several approaches simultaneously, that is, to use a multi-model approach~\cite{kulyabov_2020_lncs_multi-model-approach_en,page_book_model-thinker_en}.
	When modeling networks, it is often difficult to propose an adequate analytical model in advance.
	In addition, experimental data are not always available.
	And in this case, it seems quite reasonable to conduct preliminary natural modeling.
	Let us review possible means of natural modeling of networks.

\section{Reproducible research}
\label{sec:reproducible}

Replication is a fundamental principle of science~\cite{baker_2016_reproducibility_en}.
There is a growing concern among scientists that too little scientific research can be replicated~\cite{goodman_2016_research-reproducibility-mean_en} (the replication crisis).
Research papers often contain inadequate details and are impossible to replicate.
Many attempts to replicate famous scientific studies fail in a wide variety of disciplines.
Replicating studies with new, independent data is expensive, rarely published in high-impact journals, and sometimes even methodologically impossible.
Computationally reproducible research is often proposed as a way to improve our ability to evaluate the validity and rigor of scientific results.
Research is reproducible when others can reproduce the results of a scientific study using only the original data, code, and documentation.

Types of reproducibility:
\begin{itemize}
\item Replication---the complete repetition of an experiment by other researchers using the same methods and data.
\item Reproducibility---the ability to obtain the same results when the original data are reanalyzed using the same procedures.
\item Repeatability---obtaining similar results when an experiment is repeated within a single study.
\end{itemize}

\subsection{Barriers to reproducibility of research}

Reasons for lack of reproducibility:
\begin{itemize}
\item complexity;
\item technological changes;
\item human errors.
\end{itemize}

\subsubsection{Complexity}

Scientific research requires specialized (and often closed) knowledge and tools that may not be available to everyone who would like to reproduce the research results.
Research requires material support (for example, high-performance computer systems).
In addition, specific software is required.
Proprietary software packages are often used in research.
But this is absolutely unacceptable.
Proprietary software can be used in engineering developments in industry.
In academic research, only open source software can be used.

\subsubsection{Technological changes}

Hardware and software change rapidly.
When old tools become obsolete, research becomes less reproducible.
A minor update in software can make the entire project less reproducible.
The versions of the software used should be documented.
Software versions can be fixed by using software containers or virtual machines.

\subsubsection{Human errors}

Details of how research was conducted are always (sic!) forgotten.
Only good documentation of the research process can protect against minor errors and careless analysis.
The description of the research process should be done in low context~\cite{hall_book_silent-language_en}.
It is almost impossible to adequately describe the use of a graphical user interface.
The research process should be described in code.

\subsection{The problem of reproducibility in modeling}

The problem of reproducibility arises in all types of modeling.
But to varying degrees.
The less formalized the study, the more aspects we cannot take into account, the greater the threat of not achieving reproducibility.
The least threat to reproducibility is in the theoretical part of the model, the greatest---in the operational.
Naturally, experimental research should attract our greatest attention.
But often setting up a real experiment is not possible (mainly for financial reasons).
Therefore, we must turn our attention to natural modeling and natural experiment.

\section{Natural modeling tools}
\label{sec:soft}

Let's consider the main tools for natural network modeling.

\subsection{Classification principles}

Let us introduce the following classification of natural network modeling tools based on functionality:
\begin{itemize}
\item multisystem modeling tools;
\item single-stack modeling tools;
\item specialized modeling tools.
\end{itemize}

\subsection{Multisystem modeling tools}

Realistic modeling requires using natural models from different vendors.
Hardware images can be obtained as operating system images.
To run such images, you need to use processor emulation tools (QEMU, VirtualBox, vmWare).

\subsubsection{GNS3}

\begin{itemize}
\item Website: \url{https://www.gns3.com/}.
\item Documentation: \url{https://docs.gns3.com/}.
\item Repository:
\begin{itemize}
\item \url{https://github.com/GNS3/gns3-gui};
\item \url{https://github.com/GNS3/gns3-server};
\item \url{https://github.com/GNS3/gns3-web-ui};
\item \url{https://github.com/GNS3/dynamips}.
\end{itemize}
\item License: GPL-3.0.
\end{itemize}

The main user interface is a graphical interface~\cite{welsh_book_gns3-guide_en}.
There is no command line interface.
Supports working with virtual machine images for different network systems.
The interface can be either a separate application with a graphical interface of the operating system (Qt), or an integrated web interface.
Uses the interface of nested virtual machines.

Several types of virtual machines are used:
\begin{itemize}
\item Dynamips: for Cisco IOS emulation;
\item QEMU: for running virtual machines;
\item Docker: container integration.
\end{itemize}
Initially, the main emulator was the Dynamips emulator.

GNS3 has an extremely low entry level (mostly due to the graphical interface).
GNS3 is quite suitable for scientific network modeling~\cite{kulyabov_2014_icumt_red-gns3_en}.
However, the problem of reproducibility of research is almost insoluble.

\subsubsection{Eve-ng}

Eve-ng (Emulated Virtual Environment New Generation).

\begin{itemize}
\item Website: \url{https://www.eve-ng.net/}
\item License: EULA (\url{https://www.eve-ng.net/index.php/documentation/eula/}).
\end{itemize}

The product is based on the UNetLab project (development stopped in 2016).
Ideologically similar to GNS3 (they would compete with each other if not for the proprietary license).
Similar functionality.
Graphical interface in the form of a web interface.

Delivery options:
\begin{itemize}
\item Community Edition, project size restrictions;
\item Professonal, offers advanced functionality;
\item Corporate, collaboration capabilities.
\end{itemize}

\subsubsection{PNetLab}

\begin{itemize}
\item Website: \url{https://pnetlab.com/pages/main}.
\item Repository: \url{https://github.com/pnetlab/pnetlab_main}.
\item Documentation: \url{https://www.pnetlab.com/pages/documentation}.
\item License: unknown.
\end{itemize}

Graphical interface in the form of a web interface.

It is a fork of Eve-ng.
The functionality is the same as Eve-ng, but without the need for payment.
Eve-ng developers expressed dissatisfaction with this project (\url{https://www.eve-ng.net/forum/viewtopic.php?f=4&t=16925}).

The project looks abandoned.

\subsubsection{Containerlab}

The project was developed by Nokia.

\begin{itemize}
\item Website: \url{https://containerlab.dev/}.
\item Repository: \url{https://github.com/srl-labs/containerlab}.
\item License: BSD-3-Clause.
\end{itemize}

Virtual network labs are built on the basis of Docker containers.
Containers are managed in a special command environment.
The topology is described in a yaml file.

The vrnetlab package (\url{https://github.com/vrnetlab/vrnetlab)} allows you to connect a virtual machine inside a container.
It is possible to combine containers and virtual machines into a single topology.
Containerlab supports working with many network operating systems, presented as virtual machine images.
You can create labs that include not only network nodes, but also everything in between: telemetry stacks, databases, test equipment nodes, network servers.

\subsubsection{Netlab}

\begin{itemize}
\item Site: \url{https://netlab.tools/}.
\item License: MIT.
\end{itemize}

A framework for automating the design, deployment, and testing of network labs using the Infrastructure as Code (IaC) concept.
Configuration description via abstract functional modules, without vendor binding.
Docker, Vagrant, KVM, Virtualbox, containerlab can be used as model providers (respectively, using all its power).

\subsection{Mono-stack modeling tools}

These systems implement a single network stack.
Most often, they are based on Linux.
This allows the simulation tool to be optimized much more than multi-system tools.

\subsubsection{Mininet}

Mininet Project developed at Stanford University.

\begin{itemize}
\item Website: \url{https://mininet.org/}.
\item Repository: \url{https://github.com/mininet/mininet}.
\item License: BSD-3-Clause.
\end{itemize}

Allows you to build simple networks.
It is possible to implement OpenFlow and SDN technologies.
OpenFlow-based controller settings can be migrated to physical equipment.
Network nodes in Mininet are processes running in a network namespace.
This approach allows you to isolate hosts on one machine from each other, but at the same time each of them has its own interface.
Python API is supported.

This tool was originally created as a specialized modeling tool.
The main purpose is to model Software-Defined Networks (SDN)~\cite{lantz_2010_network-laptop-sdn_en,lantz_2015_mininet-distributed-sdn_en,oliveira_2014_mininet-emulation-sdn_en}.
But gradually there was a drift towards a general-purpose modeling package.
It seems that this happened due to the introduction of Mininet in training~\cite{ryll_2014_measuring-tcp-tail-loss_en}.
In addition, in the context of Mininet modeling, researchers have begun to pay increased attention to the reproducibility of research~\cite{handigol_2012_reproducible-network-experiments-container_en,heller_2013_reproducible-network-emulation_en,yan_2017_networking-reproducing-research_en}.

\subsubsection{Netkit}

Website: \url{https://www.netkit.org/}.

Developed for scientific research and education~\cite{pizzonia_2014_netkit-network-emulation-education_en,ariyanto_2018_teaching-network-security-netkit_en}.
The implementation was based on the User-Mode Linux (UML) subsystem.
Now has purely historical significance.
Replaced by Kathará.

\subsubsection{Kathará}

Kathará (from the Greek purely).

\begin{itemize}
\item Website: \url{https://www.kathara.org/}.
\item Repository: \url{https://github.com/KatharaFramework/Kathara}.
\item License: GPL-3.0.
\end{itemize}

It is an evolution of the Netkit network emulator~\cite{scazzariello_2020_kathara-network-emulation_en}.
In fact, it can be perceived as Netkit, in which User-Mode Linux is replaced with Docker.
Kathará allows you to simulate virtual networks based on Docker containers or Kubernetes clusters~\cite{alberro_2022_experimentation-environments-routing_en,bonofiglio_2018_kathara-container-based-framework_en}.
Supports SDN, NFV (Network function virtualization), BGP, OSPF technologies.

Declarative description language and command line tools.
There is a graphical interface: Netkit-Lab-Generator (\url{https://github.com/KatharaFramework/Netkit-Lab-Generator}), but it has a subordinate meaning.
The peculiarity of this project is its rapid revolutionary development.
Variants of this emulator constantly appear, aimed at solving new problems.
Later, most often, these new branches merge into the original project.

Let's give examples of some of them.

\paragraph{Megalos и Sybil}

The Megalos project is aimed at modeling large heterogeneous networks~\cite{scazzariello_2020_megalos-scalable-architecture-virtualization_en,scazzariello_2021_megalos-scalable-architecture-virtualization-large_en}.
The main direction of modeling is the network architecture of data centers.
VXLAN networks with BGP routing are used.

The Kubernetes orchestrator has been introduced into the Docker subsystem, which allows modeling large networks.
Distributing emulation across multiple nodes allows Megalos to emulate large-scale network infrastructures that require a significant number of virtual local area networks.

The Sibyl software platform~\cite{caiazzi_2022_sibyl-fat-trees_en} was created based on Kathará and Megalos.
Sibyl is used to model routing protocols in fat-tree networks.

\paragraph{Nested containers}

The goal of the project was to build a digital twin of a data center.
To accurately emulate a hierarchy consisting of physical servers, virtual machines, and containers, support for nested virtualization is a fundamental requirement.
To solve this problem, the ability to use nested containers was developed~\cite{caiazzi_2023_nesting-containers-emulations_en}.

\subsubsection{IMUNES}

IMUNES (Integrated Multi-protocol Network Emulator/Simulator).

\begin{itemize}
\item Website: \url{https://imunes.net/}
\item Repository: \url{https://github.com/imunes/imunes}
\item License: CC-BY 4.0.
\end{itemize}

It is possible to use both a graphical interface and a scripting language.
but the primary one is the command line interface.
Implements the FreeBSD network stack~\cite{zec_2003_clonable-network-stack-freebsd_en}.
Can be used on Linux, FreeBSD hosts.

Virtualization~\cite{salopek_2014_network-testbed-imunes_en}:
\begin{itemize}
\item FreeBSD: jails;
\item Linux: Docker.
\end{itemize}

The development is uneven.
At one time, it seemed that the project was abandoned.
This is a typical local project for specific purposes of developers.
If they do not need something, it will not be implemented.

\subsubsection{CORE}

CORE (Common Open Research Emulator).

\begin{itemize}
\item Сайт: \url{https://coreemu.github.io/core/}.
\item Репозиторий: \url{https://github.com/coreemu/core}.
\item Документация: \url{https://coreemu.github.io/core/index.html}.
\item Лицензия: BSD-2-Clause.
\end{itemize}

Virtualization: Linux namespaces.
Implements the FreeBSD~\cite{ahrenholz_2008_core-network-emulator_en} network stack.
Supports Docker containers for deploying services (DNS, HTTP).
Connections are implemented via Linux bridges, VLAN, VPN.
Dynamic routing: Quagga/FRR.
The graphical interface is used only for building the topology.

\subsubsection{Cloonix}

\begin{itemize}
\item Website: \url{https://clownix.net/}.
\item Repository: \url{https://github.com/clownix/cloonix}.
\item License: AGPLv3.
\end{itemize}

Simulation based on virtual machines and containers~\cite{linkletter_2016_open-source-network-emulators_en}.
For virtual machines: kvm.
For containers: podman.
Only Linux host is supported.
In addition to the command interface, there is a graphical interface.
Links between nodes are based on Open vSwitch.
Theoretically, the tool can support different network equipment (for example, Cisco).
However, in reality, only Linux is used.

\subsection{Specialized modeling tools}

These modeling tools are aimed at solving narrow specific problems.
This does not mean that it will be impossible to conduct complex modeling.
But it will be clearly more difficult than on universal systems.

\subsubsection{Toxiproxy}

\begin{itemize}
\item Repository: \url{https://github.com/Shopify/toxiproxy}.
\item License: MIT.
\end{itemize}

Allows you to study the fault tolerance of solutions in a network environment.
You can simulate anomalies and failures: delays in receiving a response from the server, changes in bandwidth.
You can run applications in an environment where all connections are tested under various network conditions.
Consists of two elements: a TCP proxy and a client interacting with it via HTTP.
Scripts are created using Ruby, Node.js, Python.
The client connects to the Toxiproxy daemon via HTTP.

\section{Results}
\label{sec:results}

Let's collect the tools for natural network modeling in a table (tab.~\ref{tab:comparison})
Let's highlight the parameters that are important to us for conducting reproducible studies.
The license type will correspond to the complexity of using the software.
Since all the systems under consideration use containers and virtual machines (or one of these), we will assume that they all support management of technological changes.
The presence of a command interface and a scripting language will help eliminate human error.

\begin{table}
\caption{Natural network modeling tools}
\label{tab:comparison}
\centering
\begin{tabular}{llll}
\hline
Program & Interface & Scripting language & License\\
\hline
GNS3 & GUI, Web & No & GPL-3.0\\
Eve-ng & Web & No & EULA\\
PNetLab & Web & No & Unknown\\
Containerlab & CLI & YAML, Ansible & BSD-3-Clause\\
Mininet & CLI & Python & BSD-3-Clause\\
Kathará & CLI, GUI & Python, YAML, shell & GPL-3.0\\
IMUNES & CLI, GUI & Python, YAML, shell & CC-BY 4.0\\
CORE & CLI, GUI & Python & BSD-2-Clause\\
Toxiproxy & CLI & Ruby, Node.js, Python & MIT\\
Cloonix & CLI, GUI & Python, shell & AGPLv3\\
netlab & CLI & YAML, Ansible, Python, shell & MIT\\
\hline
\end{tabular}
\end{table}

\section{Discussion}
\label{sec:discussion}

First of all, tools with non-free licenses should be removed from consideration: Eve-ng and PNetLab.

GNS3 does not support automation via code.
The lack of a command line interface prevents reproducible studies.
However, GNS3 is well suited for teaching networking and the principles of real-world modeling.

The other modeling tools we have reviewed are quite suitable for implementing reproducible studies.

For modeling heterogeneous networks consisting of equipment from different manufacturers, it remains to recommend containerlab and netlab.
If it is necessary to model not so much the protocols themselves, but the features of their implementation in equipment from different manufacturers, there is simply no alternative.
However, nothing prevents you from using these tools simply for modeling arbitrary networks.

Kathará supports distributed execution, making it indispensable for resource-intensive modeling.
In addition, the tool explicitly addressed the methodology for creating digital twins of networked systems.

Mininet is developing extremely slowly.
But this environment is very easy to use.
In addition, this modeling tool is included in the educational program of many universities.
And a researcher, all other things being equal, will prefer to use the modeling tool with which he is familiar.

Toxiproxy is designed to study network security issues.
Unlike other modeling tools, we have not studied it in depth.
However, modeling network security is also within our interests.

The IMUNES, CORE, Cloonix modeling tools look quite decent.
But, the winner takes all.
These modeling tools do not offer, at the moment, any special advantages over the systems we have highlighted.

\section{Conclusion}
\label{sec:conclusion}

Let's summarize our recommendations.
\begin{itemize}
\item For initial training of students---GNS3.
\item For modeling simple systems and teaching natural modeling---Mininet.
\item For modeling complex systems, resource-intensive modeling---Kathará, containerlab, netlab.
\item For modeling heterogeneous networks, maximum correspondence to real systems---containerlab, netlab.
\end{itemize}

This work marks a milestone in our research.
We began our field-based network studies with GNS3.
The reason we abandoned this tool was the impossibility of conducting reproducible research within it.
This naturally led us to Mininet.
But very quickly, the baseline tool became a hindrance rather than a help.
Therefore, following this study, we are focusing on Kathará and containerlab.
Our next step is to conduct several studies, implementing field experiments in these three environments.
To practically confirm (or refute) our findings.

\begin{acknowledgments}
This paper has been
financially supported within the framework of the RUDN University
Development Program ``Strategy-2030,''
Project C05.1 ``Implementation of international and domestic academic cooperation and enhancement of academic reputation.''

\end{acknowledgments}

\printbibliography

\end{document}


\graphicspath{{image/reproducible-network-research/ru/}{image/reproducible-network-research/}{image/}}

\title{Воспроизводимые исследования при моделировании сетей}
\author{А. М. Ермолаева}
\email{ermolaevaanna@bk.ru}
\affiliation{Российский университет дружбы народов им. Патриса Лумумбы, ул. Миклухо-Маклая, д. 6, Москва, 117198, Российская Федерация}

\author{Т. Р. Велиева}
\email{velieva-tr@rudn.ru}
\affiliation{Российский университет дружбы народов им. Патриса Лумумбы, ул. Миклухо-Маклая, д. 6, Москва, 117198, Российская Федерация}

\author{А. А. Живцова}
\email{zhivtsova-aa@rudn.ru}
\affiliation{Российский университет дружбы народов им. Патриса Лумумбы, ул. Миклухо-Маклая, д. 6, Москва, 117198, Российская Федерация}

\author{А. В. Королькова}
\email{korolkova-av@rudn.ru}
\affiliation{Российский университет дружбы народов им. Патриса Лумумбы, ул. Миклухо-Маклая, д. 6, Москва, 117198, Российская Федерация}

\author{Д. С. Кулябов}
\email{kulyabov-ds@rudn.ru}
\affiliation{Российский университет дружбы народов им. Патриса Лумумбы, ул. Миклухо-Маклая, д. 6, Москва, 117198, Российская Федерация}
\affiliation{Объединённый институт ядерных исследований, ул. Жолио-Кюри, д. 6, Дубна, 141980, Российская Федерация}

    \begin{abstract}
        \emph{Предпосылки}
        При моделировании сетей возникает проблема получения экспериментальных данных для верификации других модельных подходов.
        И даже при наличии каких-либо экспериментальных данных необходимо быть уверенным в их достоверности.
        \emph{Цель}
        Необходимо предложить методы получения достоверных экспериментальных данных.
        \emph{Методы}
        По своей природе сетевое оборудование есть программно-аппаратный комплекс.
        Поэтому натурная программная модель может считаться полностью эквивалентной реальному оборудования.
        И реальный эксперимент можно заменить на натурный эксперимент.
        Достоверность же натурного эксперимента будет базироваться на его воспроизводимости.
        \emph{Результаты}
        Проведено сравнение популярных пакетов натурного сетевого моделирование.
        Проведено разделение этих пакетов по функциональности и реализуемости воспроизводимых исследований.
        \emph{Выводы}
        Большинство программных пакетов удовлетворяет критериям воспроизводимости.
        Выбор конкретного решения зависит от нетехнических факторов: популярности и знания пакета.
    \end{abstract}

    \keywords{%
        натурное моделирование,
        воспроизводимое исследование,
        исследование как код
    }

        \maketitle

    \section{Введение}
\label{sec:intro}

    При моделировании сетей возникает проблема доступа к экспериментальным данным.
    Редко когда есть возможность произвести эксперимент на реальном оборудовании.
    Без экспериментальных данным мы теряем возможность верифицировать результаты разным типов моделирования.
    Для снятия проблемы можно рекомендовать проведение натурного эксперимента с помощью средства натурного моделирования.
    Следует заметить, натурное моделирование компьютерных систем выделяется тем, что компьютерные системы можно точно представлять с помощью других компьютерных систем.
    Однако при этом возникает проблема воспроизводимости проведения натурного эксперимента.
    Целью статьи является выбор средства натурного моделирования, когда во главу угла ставится именно поддержка этим средством проведения воспроизводимого эксперимента.

    \subsection{Структура статьи}
\label{sec:structure}

    В разделе \ref{sec:model} даётся краткий обзор методов моделирования.
    В разделе \ref{sec:reproducible} приводятся рассуждения о проблемах, мешающим проводить воспроизводимые исследования.
    В разделе \ref{sec:soft} приводится обзор средств натурного моделирования сетей.
    Авторы опробовали эти средства моделирования и описали свои впечатления от них.
    В разделе \ref{sec:results} суммированы характеристики рассмотренных пакетов, относящиеся к воспроизводимым вычислениям.
    В разделе \ref{sec:discussion} приведено обсуждение, какие пакеты наиболее подходят для воспроизводимого натурного моделирования.
    В разделе \ref{sec:conclusion} приведены конкретные наши рекомендации по использованию рассмотренных пакетов моделирования.

	\section{Подходы к моделированию}
\label{sec:model}

	Моделирование как дисциплина охватывает разные типы модельных подходов~\cite{kulyabov_2024_dcm_editorial_journal-rubrics_en}.
	С нашей точки зрения эти подходы можно схематически описать единым образом.
	Исследовательская структура состоит из операциональных и теоретических частей (рис.~\ref{fig:model-generic}).
	Операциональные части представлены процедурами приготовления системы и измерения. Также возможно описание операциональных частей как входных и выходных данных.
	Теоретическая часть состоит из двух слоёв: модельного слоя и слоя реализации.

\begin{figure}
  \centering
  \includegraphics[width=0.6\linewidth]{model-generic}
		\caption{Общая структура модельного подхода}
  \label{fig:model-generic}
\end{figure}

	Модельный слой является основным и задаёт собственно исследуемую модель.
	Слой реализации описывает конкретную структуру эволюции системы. В зависимости от типа реализации, можно получать разные виды моделей:
	\begin{itemize}
		\item реализация --- математические выражения: аналитические математические модели;

		\item реализация --- аналоговая система: физическая модель;

		\item реализация --- эквивалентная система: натурная модель;

		\item реализация --- алгоритм: имитационные модели;

		\item реализация --- аппроксимация поведения: суррогатная модель.
	\end{itemize}

	Каждый тип моделей имеет свою область применимости, свои преимущества и недостатки.
	Использование всего спектра моделей позволяет наиболее глубокое и всестороннее исследование моделируемой системы.

	\subsection{Аналитическое моделирование}

	Наиболее строгое исследование базируется, обычно, на аналитической математической модели.
	В этом случае модельный слой реализуется посредством математических выражений, описывающих эволюцию системы.

	\subsection{Физическое моделирование}

	Полученную математическую модель нужно сопоставить с экспериментальными данными, верифицировать её.
	Для этого можно создать модель, аналогичную физической или технической системе.
	Исследование проводится на установках, сохраняющих природу явлений, но не обязательно идентичных реальному объекту.
	Модель может быть уменьшенной или увеличенной копией прототипа, где критически важно соблюдение критериев подобия.
	В основе физического моделирования лежат теория подобия и анализ размерностей.

	\subsection{Натурное моделирование}

	Эксперименты проводятся на реальном объекте или его части в условиях, максимально приближённых к эксплуатационным.
	Включает производственные испытания, научные эксперименты с вмешательством в процесс и анализ накопленного опыта.
	Натурная модель может быть и виртуальной.
	Например, можно построить модель сети передачи данных, используя образы операционных систем маршрутизаторов и коммутаторов.
	Использование виртуальной модели радикально упрощает процесс моделирования.

	\subsection{Имитационное моделирование}

	С развитием компьютерной техники появилась возможность задавать модельную реализацию не в виде математического описания, а в виде некоторого алгоритма.
	Такого типа модели получили название имитационных моделей, а сам подход получил название имитационного моделирования.
	Имитационная модель играет двоякую роль.
	Отлаженная и проверенная на экспериментальных данных, физической и натурной моделях имитационная модель сама по себе может служить целям верификации математической модели.
	С другой стороны, имитационная модель позволяет более эффективно, чем математическая модель, исследовать поведение моделируемой системы при разных вариантах входных данных.

	\subsection{Статистическое моделирование}

	Это совершенно отдельное направление, использующее для автоматического построения моделей не каузацию, а корреляцию.
	К данному типу моделирования относят модели, которые реализуются посредством методов машинного обучения.
	Его можно разделить на несколько подходов.
	Для целей собственно моделирования наиболее интересно суррогатное моделирование.
	В этом подходе модельный слой известен и даже имеет реализацию (чаще всего в виде аналитической или имитационной модели).
	Исходное моделирование может занять достаточно продолжительное время.
	Для упрощения исследования строятся модели, которые имитируют поведение исходной модели настолько близко, насколько это возможно, в то время как вычислительно дёшевы.
	Суррогатные модели строятся с использованием подхода, основанного на данных. Точная внутренняя работа кода моделирования не предполагается известной (или даже понятой), важно только поведение ввода–вывода (приготовления–измерения).
	Научная задача суррогатного моделирования заключается в создании суррогата, который является максимально точным, используя как можно меньше оценок моделирования~\cite{kulyabov_2019_ceur-ws_2507_structural-deep-learning_en}.

	\subsection{Мультимодельный подход}

	При моделировании естественно использовать одновременно несколько подходов, то есть использовать мультимодельный подход~\cite{kulyabov_2020_lncs_multi-model-approach_en,page_book_model-thinker_en}.
	При моделировании сетей зачастую трудно заранее предложить адекватную аналитическую модель.
	Кроме того, не всегда в доступе есть экспериментальные данные.
	И в данном случае представляется вполне обоснованным провести предварительно натурное моделирование.
	Сделаем обзор возможных средств натурного моделирования сетей.

\section{Воспроизводимые исследования}
\label{sec:reproducible}

Воспроизводимость является фундаментальным принципом науки~\cite{baker_2016_reproducibility_en}.
Среди учёных растёт опасение, что слишком мало научных исследований может быть воспроизведено~\cite{goodman_2016_research-reproducibility-mean_en} (кризис репликации).
Научные статьи часто содержат неадекватные детали, их невозможно воспроизвести.
Многие попытки воспроизвести известные научные исследования в самых разных дисциплинах терпят неудачу.
Повторение исследований с новыми независимыми данными стоит дорого, редко публикуется в авторитетных журналах, а иногда даже методологически невозможно.
Воспроизводимое с помощью вычислений исследование часто предлагается в качестве пути повышения нашей способности оценивать обоснованность и строгость научных результатов.
Исследование воспроизводимо, когда другие могут воспроизвести результаты научного исследования, используя только исходные данные, код и документацию.

Виды воспроизводимости:
\begin{itemize}
\item Репликация (replication) --- полное повторение эксперимента другими исследователями с использованием тех же методов и данных.
\item Воспроизводимость (reproducibility) --- способность получить те же результаты при повторном анализе исходных данных с помощью тех же процедур.
\item Повторяемость (repeatability) --- получение схожих результатов при повторении эксперимента в рамках одного исследования.
\end{itemize}

\subsection{Барьеры для воспроизводимости исследования}

Причины отсутствия воспроизводимости:
\begin{itemize}
\item сложность;
\item технологические изменения;
\item человеческие ошибки.
\end{itemize}

\subsubsection{Сложность}

Научные исследования требуют специальных (и часто закрытых) знаний и инструментов, которые могут быть доступны не всем, кто хотел бы воспроизвести результаты исследований.
Для исследования необходимо материальное обеспечение (например, высокопроизводительные компьютерные системы).
Кроме того, требуются специфическое программное обеспечение.
При проведении исследований часто используют проприетарные программные пакеты.
Но это категорически недопустимо.
Проприетарное программное обеспечение может использоваться в инженерных разработках в индустрии.
В академических исследования возможно применение только открытого программного обеспечения.

\subsubsection{Технологические изменения}

Аппаратное и программное обеспечение меняются быстро.
Когда старые инструменты устаревают, исследования становятся менее воспроизводимыми.
Незначительное обновление в программном обеспечении может сделать весь проект менее воспроизводимым.
Следует документирование версий используемого программного обеспечения.
Можно фиксировать версию программного обеспечения посредством использования программных контейнеров или виртуальных машин.

\subsubsection{Человеческая ошибка}

Детали того, как проводились исследования, всегда (sic!) забываются.
Только хорошее документирование процесс исследования может защитить от мелких ошибок и небрежного анализа.
Описание процесса исследования должно выполняться в низком контексте~\cite{hall_book_silent-language_en}.
Практически невозможно адекватно описать использование графического интерфейса пользователя.
Следует стремиться к тому, чтобы процесс исследования описывался в виде кода.

\subsection{Проблема воспроизводимости в моделировании}

Проблема воспроизводимости возникает при всех типах моделирования.
Но в разной степени.
Чем менее формализовано исследование, чем больше аспектов, которые мы не можем учесть, тем больше угроза не достичь воспроизводимости.
Наименьшая угроза воспроизводимости находится в теоретической части модели, наибольшая --- в операциональной.
Естественно, экспериментальное исследование должно привлекать наибольшее наше внимание.
Но зачастую постановка реально эксперимента не представляется возможной (в основном по финансовым сообранониям).
Поэтому мы должны обратить своё внимание к натурному моделированию и натурному эксперименту.

\section{Средства натурного моделирования}
\label{sec:soft}

Рассмотрим основные средства натурного моделирования сетей.

\subsection{Принципы классификации}

Введём следующую классификацию средств натурного моделирования сетей на основе функциональности:
\begin{itemize}
\item мультисистемные средства моделирования;
\item моностековые средства моделирования;
\item специализированные средства моделирования.
\end{itemize}

\subsection{Мультисистемные средства моделирования}

Для реалистичного моделирования требуется использовать натурные модели разных вендоров.
Образы оборудования можно получить в виде образов операционной системы.
Для запуска таких образов требуется использование средств эмуляции процессора (QEMU, VirtualBox, vmWare).

\subsubsection{GNS3}

\begin{itemize}
\item Сайт: \url{https://www.gns3.com/}.
\item Документация: \url{https://docs.gns3.com/}.
\item Репозиторий:
\begin{itemize}
\item \url{https://github.com/GNS3/gns3-gui};
\item \url{https://github.com/GNS3/gns3-server};
\item \url{https://github.com/GNS3/gns3-web-ui};
\item \url{https://github.com/GNS3/dynamips}.
\end{itemize}
\item Лицензия: GPL-3.0.
\end{itemize}

Основной интерфейс пользователя --- графический интерфейс~\cite{welsh_book_gns3-guide_en}.
Интерфейс командной строки отсутствует.
Поддерживает работу с образами виртуальных машин для разных сетевых систем.
Интерфейс может быть как в виде отдельного приложения с графическим интерфейсом операционной системы (Qt), так и в виде встроенного web-интерфейса.
Использует интерфейс вложенных виртуальных машин.

Используется несколько типов виртуальных машин:
\begin{itemize}
\item Dynamips: для эмуляции Cisco IOS;
\item QEMU: для запуска виртуальных машин;
\item Docker: интеграция контейнеров.
\end{itemize}
Изначально основным эмулятором был эмулятор Dynamips.

GNS3 имеет крайне низкий уровень входа (в основном из-за графического интерфейса).
GNS3 вполне применим для научного моделирования сетей~\cite{kulyabov_2014_icumt_red-gns3_en}.
Однако проблема воспроизводимости исследований практически неразрешима.

\subsubsection{Eve-ng}

Eve-ng (Emulated Virtual Environment New Generation).

\begin{itemize}
\item Сайт: \url{https://www.eve-ng.net/}
\item Лицензия: EULA (\url{https://www.eve-ng.net/index.php/documentation/eula/}).
\end{itemize}

Продукт основан на проекте UNetLab (разработка прекращена в 2016 году).
Идеологически похож на GNS3 (они бы и конкурировали между собой, если бы не проприетарная лицензия).
Сходный функционал.
Графический интерфейс в виде web-интерфейса.

Варианты поставки:
\begin{itemize}
\item Community Edition, ограничения по размерам проекта;
\item Professonal, предлагает расширенную функциональность;
\item Corporate, возможность совместной работы.
\end{itemize}

\subsubsection{PNetLab}

\begin{itemize}
\item Сайт: \url{https://pnetlab.com/pages/main}.
\item Репозиторий: \url{https://github.com/pnetlab/pnetlab_main}.
\item Документация: \url{https://www.pnetlab.com/pages/documentation}.
\item Лицензия: не известно.
\end{itemize}

Графический интерфейс в виде web-интерфейса.

Является форком Eve-ng.
Функциональность такая-же, как у Eve-ng, но без необходимости оплаты.
Разработчики Eve-ng высказывали неудовольствие этим проектом (\url{https://www.eve-ng.net/forum/viewtopic.php?f=4&t=16925}).

Проект выглядит заброшенным.

\subsubsection{Containerlab}

Проект разработан Nokia.

\begin{itemize}
\item Сайт: \url{https://containerlab.dev/}.
\item Репозиторий: \url{https://github.com/srl-labs/containerlab}.
\item Лицензия: BSD-3-Clause.
\end{itemize}

Виртуальные сетевые лаборатории строятся на базе контейнеров Docker.
Управление контейнерами происходит в специальном командном окружении.
Топология описывается в yaml-файле.

Пакет vrnetlab (\url{https://github.com/vrnetlab/vrnetlab)} позволяет подключать виртуальную машину внутрь контейнера.
Возможно объединять контейнеры и виртуальные машины в единую топологию.
Containerlab поддерживает работу со множеством сетевых операционных систем, представленных в виде образов виртуальных машин.
Можно создавать лаборатории, которые включают не только сетевые узлы, но и все, что между ними: стеки телеметрии, базы данных, узлы тестового оборудования, сетевые серверы.

\subsubsection{Netlab}

\begin{itemize}
\item Сайт: \url{https://netlab.tools/}.
\item Лицензия: MIT.
\end{itemize}

Фреймворк для автоматизации проектирования, развертывания и тестирования сетевых лабораторий с использованием концепции Infrastructure as Code (IaC).
Описание конфигураций через абстрактные функциональные модули, без привязки к вендору.
В качестве модельных провайдеров можно использовать Docker, Vagrant, KVM, Virtualbox, containerlab (соответственно, используя всю его мощь).

\subsection{Моностековые средства моделирования}

Данные системы реализуют какой-либо один сетевой стек.
Чаще всего на основе Linux.
Это позволяет оптимизировать средство моделирования намного сильнее, чем мультисистемные средства.

\subsubsection{Mininet}

Проект Mininet разработан в Стэнфордском университете.

\begin{itemize}
\item Сайт: \url{https://mininet.org/}.
\item Репозиторий: \url{https://github.com/mininet/mininet}.
\item Лицензия: BSD-3-Clause.
\end{itemize}

Позволяет строить простые сети.
Можно реализовать технологии OpenFlow и SDN.
Настройки контроллеров на базе OpenFlow можно мигрировать на физическое оборудование.
Узлы сети в Mininet представляют собой процессы, запущенные в сетевом пространстве имён.
Такой подход позволяет изолировать хосты на одной машине друг от друга, но при этом каждый из них имеет собственный интерфейс.
Поддерживается Python API.

Данное средство первоначально было создано как специализированное средство моделирования.
Основное предназначение --- моделирование Software-Defined Networks (SDN)~\cite{lantz_2010_network-laptop-sdn_en,lantz_2015_mininet-distributed-sdn_en,oliveira_2014_mininet-emulation-sdn_en}.
Но постепенно произошёл дрейф в сторону универсального пакета моделирования.
Представляется, что это произошло из-за внедрения Mininet в обучение~\cite{ryll_2014_measuring-tcp-tail-loss_en}.
Кроме того, в рамках моделирования на Mininet исследователи стали уделять повышенное внимание воспроизводимости исследований~\cite{handigol_2012_reproducible-network-experiments-container_en,heller_2013_reproducible-network-emulation_en,yan_2017_networking-reproducing-research_en}.

\subsubsection{Netkit}

Сайт: \url{https://www.netkit.org/}.

Разрабатывался для научных исследований и обучения~\cite{pizzonia_2014_netkit-network-emulation-education_en,ariyanto_2018_teaching-network-security-netkit_en}.
Реализация строилась на подсистеме User-Mode Linux (UML).
Сейчас имеет исключительно историческое значение.
Заменён на Kathará.

\subsubsection{Kathará}

Kathará (от греческого чистый).

\begin{itemize}
\item Сайт: \url{https://www.kathara.org/}.
\item Репозиторий: \url{https://github.com/KatharaFramework/Kathara}.
\item Лицензия: GPL-3.0.
\end{itemize}

Является развитием сетевого эмулятора Netkit~\cite{scazzariello_2020_kathara-network-emulation_en}.
Фактически можно воспринимать как Netkit, в котором User-Mode Linux заменён на Docker.
Kathará позволяет моделировать виртуальные сети на базе контейнеров Docker или кластеров Kubernetes~\cite{alberro_2022_experimentation-environments-routing_en,bonofiglio_2018_kathara-container-based-framework_en}.
Поддерживает технологии SDN, NFV (Network function virtualization), BGP, OSPF.

Декларативный язык описания и средства командной строки.
Есть графический интерфейс: Netkit-Lab-Generator (\url{https://github.com/KatharaFramework/Netkit-Lab-Generator}), но он имеет подчинённое значение.
Особенностью данного проекта является его быстрое революционное развитие.
Постоянно появляются варианты данного эмулятора, направленные на решение новых задач.
В дальнейшем, чаще всего, эти новые ответвления вливаются в исходный проект.

Приведём примеры некоторых из них.

\paragraph{Megalos и Sybil}

Проект Megalos направлен на моделирование больших гетерогенных сетей~\cite{scazzariello_2020_megalos-scalable-architecture-virtualization_en,scazzariello_2021_megalos-scalable-architecture-virtualization-large_en}.
Основное направление моделирования --- сетевая архитектура центров обработки данных.
Используются сети VXLAN с маршрутизацией BGP.

В подсистему Docker введён оркестратор Kubernetes, что позволяет моделировать большие сети.
Распределение эмуляции по нескольким узлам позволяет Megalos эмулировать крупномасштабные сетевые инфраструктуры, где требуется значительное количество виртуальных локальных сетей.

На основе Kathará и Megalos была создана программная платформа Sibyl~\cite{caiazzi_2022_sibyl-fat-trees_en}.
Sibyl используется для моделирование протоколов маршрутизации в сетях с толстыми деревьями.

\paragraph{Вложенные контейнеры}

Задача проекта состояла в построении цифрового двойника центра обработки данных.
Для точной эмуляции иерархии, состоящей из физических серверов, виртуальных машин и контейнеров, поддержка вложенной виртуализации является фундаментальным требованием.
Для решения этой задачи и была разработана возможность использования вложенных контейнеров~\cite{caiazzi_2023_nesting-containers-emulations_en}.

\subsubsection{IMUNES}

IMUNES (Integrated Multi-protocol Network Emulator/Simulator).

\begin{itemize}
\item Сайт: \url{https://imunes.net/}
\item Репозиторий: \url{https://github.com/imunes/imunes}
\item Лицензия: CC-BY 4.0.
\end{itemize}

Возможно использование как графического интерфейса, так и скриптового языка.
но первичным является интерфейс командной строки.
Реализует сетевой стек FreeBSD~\cite{zec_2003_clonable-network-stack-freebsd_en}.
Можно использовать на хостах Linux, FreeBSD.

Виртуализация~\cite{salopek_2014_network-testbed-imunes_en}:
\begin{itemize}
\item FreeBSD: jails;
\item Linux: Docker.
\end{itemize}

Развитие неравномерное.
Одно время создавалось впечатление, что проект заброшен.
Это типичный локальный проект для конкретных целей разработчиков.
Если что не нужно им, то это и не будет реализовано.

\subsubsection{CORE}

CORE (Common Open Research Emulator).

\begin{itemize}
\item Сайт: \url{https://coreemu.github.io/core/}.
\item Репозиторий: \url{https://github.com/coreemu/core}.
\item Документация: \url{https://coreemu.github.io/core/index.html}.
\item Лицензия: BSD-2-Clause.
\end{itemize}

Виртуализация: Linux namespaces.
Реализует сетевой стек FreeBSD~\cite{ahrenholz_2008_core-network-emulator_en}.
Поддерживает Docker-контейнеры для развертывания сервисов (DNS, HTTP).
Связи реализуются через Linux-мосты, VLAN, VPN.
Динамическая маршрутизация: Quagga/FRR.
Графический интерфейс используется только для построения топологии.

\subsubsection{Cloonix}

\begin{itemize}
\item Сайт: \url{https://clownix.net/}.
\item Репозиторий: \url{https://github.com/clownix/cloonix}.
\item Лицензия: AGPLv3.
\end{itemize}

Моделирование на базе виртуальных машин и контейнеров~\cite{linkletter_2016_open-source-network-emulators_en}.
Для виртуальных машин: kvm.
Для контейнеров: podman.
Поддерживается работа только на Linux.
Кроме командного интерфейса есть графический интерфейс.
Линки между узлами основаны на Open vSwitch.
Теоретически средство может поддерживать разное сетевой оборудование (например, Cisco).
Однако реально используется только Linux.

\subsection{Специализированные средства моделирования}

Данные средства моделирования направлены на решение узких конкретных задач.
Это вовсе не означает, что нельзя будет проводить комплексное моделирование.
Но это будет явно сложнее, чем на универсальных системах.

\subsubsection{Toxiproxy}

\begin{itemize}
\item Репозиторий: \url{https://github.com/Shopify/toxiproxy}.
\item Лицензия: MIT.
\end{itemize}

Позволяет изучить отказоустойчивость решений в сетевой среде.
Можно имитировать аномалии и сбои: задержки при получении ответа от сервера, изменение пропускной способность.
Можно запускать приложения в среде, где все соединения проходят проверку в различных сетевых условиях.
Состоит из двух элементов: TCP-прокси и клиента, взаимодействующего с ним по HTTP.
Скрипты  создаются с помощью Ruby, Node.js, Python.
Клиент подключается к демону Toxiproxy по HTTP.

\section{Результаты}
\label{sec:results}

Соберём средства натурного моделирования сетей в таблицу (таб.~\ref{tab:comparison})
Выделим те параметры, которые важны нам для проведения воспроизводимых исследований.
Тип лицензии будет соответствовать сложности использования программного обеспечения.
Поскольку все рассматриваемые системы используют контейнеры и виртуальные машины (либо что-то одно из этого), то будем считать, что все они поддерживают управление технологическими изменениями.
Наличие командного интерфейса и скриптового языка будет способствовать устранению человеческой ошибки.

\begin{table}
  \caption{Средства натурного моделирования сетей}
  \label{tab:comparison}
\centering
\begin{tabular}{llll}
\hline
Программа & Интерфейс & Скриптовый язык & Лицензия\\
\hline
GNS3 & GUI, Web & Нет & GPL-3.0\\
Eve-ng & Web & Нет & EULA\\
PNetLab & Web & Нет & Не известно\\
Containerlab & CLI & YAML, Ansible & BSD-3-Clause\\
Mininet & CLI & Python & BSD-3-Clause\\
Kathará & CLI, GUI & Python, YAML, shell & GPL-3.0\\
IMUNES & CLI, GUI & Python, YAML, shell & CC-BY 4.0\\
CORE & CLI, GUI & Python & BSD-2-Clause\\
Toxiproxy & CLI & Ruby, Node.js, Python & MIT\\
Cloonix & CLI, GUI & Python, shell & AGPLv3\\
netlab & CLI & YAML, Ansible, Python, shell & MIT\\
\hline
\end{tabular}
\end{table}

\section{Обсуждение}
\label{sec:discussion}

В первую очередь следует удалить из рассмотрения инструментарий с несвободными лицензиями: Eve-ng и PNetLab.

GNS3 не поддерживает возможность автоматизации с помощью кода.
Отсутствие интерфейса командной строки не позволяет проводить воспроизводимые исследования.
Впрочем, GNS3 хорошо подходит для обучения работы с сетями и принципам проведения натурного моделирования.

Остальные рассмотренные нами средства моделирования вполне подходят для реализации воспроизводимых исследований.

Для моделирования гетерогенных сетей, состоящий из оборудования разных производителей, остаётся рекомендовать containerlab и netlab.
Если необходимо моделировать не столько протоколы сами по себе, а особенности их реализации в оборудовании разных производителей, альтернативы просто нет.
Впрочем, ничего не мешает использовать эти средства и просто для моделирования произвольных сетей.

Kathará поддерживает распределённое выполнение, поэтому незаменима для ресурсоёмкого моделирования.
Кроме того, в рамках данного средства явно рассматривалась методика создания цифровых двойников сетевых систем.

Mininet развивается крайне неторопливо.
Но эта среда кране проста в применении.
Кроме того, это средство моделирования входит в образовательную программу многих вузов.
А исследователь, при прочих равных условиях, предпочтёт использовать то средство моделирования, с которым он знаком.

Toxiproxy предназначен для исследования вопросов сетевой безопасности.
В отличии от других средств моделирования мы не исследовали его глубоко.
Впрочем, моделирование сетевой безопасности тоже входит в круг наших интересов.

Средства моделирования IMUNES, CORE, Cloonix выглядят вполне достойными.
Но, победитель получает всё.
Эти средства моделирования не предлагают, на данный момент, никаких особенных преимуществ относительно выделенных нами систем.

\section{Заключение}
\label{sec:conclusion}

Резюмируем наши рекомендации.
\begin{itemize}
\item Для первичного обучения студентов --- GNS3.
\item Для моделирования простых систем и обучения натурному моделированию --- Mininet.
\item Для моделирования сложных систем, ресурсоёмкого моделирования --- Kathará, containerlab, netlab.
\item Для моделирования гетерогенных сетей, максимального соответствия реальным системам --- containerlab, netlab.
\end{itemize}

  Данная работа является определённой вехой в наших исследованиях.
  Мы начинали натурные сетевые исследования с GNS3.
  Причиной, побудившей нас отойти от этого средства, стала невозможность осуществления воспроизводимых исследований в его рамках.
  Что естественным образом привело нас к Mininet.
  Но очень скоро донное средство стало тормозом, а не помощником.
  Поэтому, после данного исследования, мы делаем ставку на Kathará и containerlab.
  Следующим шагом мы предполагаем провести несколько исследований, с реализацией натурного эксперимента в этих трёх средах.
  Чтобы практически подтвердить (или опровергнуть) наши выводы.

\def\acknowledgmentsname{Благодарности}

\begin{acknowledgments}
Публикация выполнена при поддержке
программы развития РУДН «Стратегия-2030»,
проект С05.1 «Реализация международного и внутрироссийского академического сотрудничества и повышение академической репутации».
\end{acknowledgments}

\printbibliography